\documentclass[a4paper]{jpconf}
\usepackage{graphicx}
\newcommand{\al}{$\alpha$}

\newcommand{\ran}{($\alpha$,$n$)}
\newcommand{\rap}{($\alpha$,$p$)}

\newcommand{\Nsv}{$N_A$$\left< \sigma v \right>$}
\newcommand{\sreac}{$\sigma_{\rm{reac}}$}
\newcommand{\sred}{$\sigma_{\rm{red}}$}
\newcommand{\Ered}{$E_{\rm{red}}$}
\newcommand{\sfact}{$S$-factor}
\begin{document}
\title{$\mathbf{\alpha}$-induced reaction cross sections in the mass range
  $\mathbf{A \approx 20 - 50}$: a critical review}

\author{Peter Mohr}

\address{
Diakonie-Klinikum, Diakoniestra{\ss}e 10, D-74523 Schw\"abisch Hall,
  Germany \\
Institute for Nuclear Research ATOMKI, P.O.\ Box 51, H-4001 Debrecen, Hungary
}

\ead{WidmaierMohr@t-online.de; mohr@atomki.mta.hu}

\begin{abstract}
In a recent review it was shown that the cross sections of \al -induced
reactions in the $A \approx 20 - 50$ mass range follow a general and smooth
trend in most cases. For comparison of cross sections of different targets at
various energies the method of reduced cross sections \sred\ and reduced
energies \Ered\ was used. Four outliers were identified: $^{36}$Ar and
$^{40}$Ar with unusally small cross sections and $^{23}$Na and $^{33}$S with
unusually large cross sections. New data for $^{23}$Na were presented at this
NPA-7 conference; contrary to the previous data, these new data fit into the
general systematics. In addition, a relation between the most effective energy
$E_0$ for astrophysical reaction rates (the so-called Gamow window) and the
reduced energy \Ered\ is presented.
\end{abstract}

\section{Introduction}
\label{sec:intro}
In a recent review \cite{Mohr15} the cross sections of \al -induced reactions
for target nuclei in the mass range $A \approx 20 - 50$ were studied. For
comparison of cross sections of different targets at various energies the
method of reduced cross sections \sred\ and reduced energies
\Ered\ \cite{Gom05} was used:
\begin{eqnarray}
E_{\rm{red}} & = & \frac{\bigl(A_P^{1/3}+A_T^{1/3}\bigr) E_{\rm{c.m.}}}{Z_P
  Z_T} 
\label{eq:Ered} \\
\sigma_{\rm{red}} & = & \frac{\sigma_{\rm{reac}}}{\bigl(A_P^{1/3}+A_T^{1/3}\bigr)^2}
\label{eq:sred}
\end{eqnarray}
The reduced energy \Ered\ takes into account the different heights of the
Coulomb barrier in the systems under consideration, whereas the reduced
reaction cross section \sred\ scales the measured total reaction cross section
\sreac\ according to the geometrical size of the projectile-plus-target
system. (All energies are given as $E_{\rm{c.m.}}$ except explicitly noted.) In
the $A \approx 20 - 50$ mass range under study the total reaction cross
section \sreac\ can often be taken from the dominating \ran\ or \rap\ reaction
channel. It was found in \cite{Mohr15} that most of the experimental cross
sections show a general and smooth trend for \sred\ {\it vs.}\ \Ered . The two
outliers $^{36}$Ar and $^{40}$Ar show significantly smaller cross sections;
however, this finding is based only on one very old experiment
\cite{Schwartz56}. Surprisingly, two recent experiments on $^{23}$Na
\cite{Alm14} and $^{33}$S \cite{Bowers13} show unexpected huge cross
sections. In addition, the data for $^{23}$Na \cite{Alm14} show a steeper
energy dependence, and the data for $^{33}$S show a flatter energy dependence
compared to most of the other nuclei in this mass region.

The present study presents two extensions to the recent review
\cite{Mohr15}. First, a relation between the reduced energy \Ered\ and the
most effective energy $E_0$ for astrophysical reaction rates (the so-called
Gamow window) is derived. Second, the review \cite{Mohr15} is updated with two
data sets for the previous outlier $^{23}$Na which were presented at this
NPA-7 conference \cite{Hub15,Tom15} and have been published now
\cite{How15PRL,Tom15PRL}. These new data for $^{23}$Na show that
$^{23}$Na should not be considered as outlier any more. Some further
information on the other outlier $^{33}$S was already given in \cite{Mohr14}.

\section{Gamow window and reduced energy \Ered }
\label{sec:gamow}
The astrophysical reaction rate \Nsv\ is defined by the Maxwellian-averaged
cross section $\left< \sigma v \right>$
\begin{equation}
\left< \sigma v \right> = \int_0^\infty \Phi(v) \, \sigma(v) \, v \, dv =
\left( \frac{8}{\pi \mu} \right)^{1/2} \frac{1}{(kT)^{3/2}} 
\int_0^\infty E \, \sigma(E) \, \exp{\left(\frac{-E}{kT}\right)} \, dE
\label{eq:rate}
\end{equation}
For an energy-independent astrophysical \sfact\ the integrand in
Eq.~(\ref{eq:rate}) shows a maximum at the energy $E_0$ which is given by
\begin{equation}
E_0 \approx 0.122\,{\rm{MeV}} \left( Z_P^2 Z_T^2 A_{\rm{red}} T_9^2
\right)^{1/3}
\label{eq:E0}
\end{equation}
with the charge numbers $Z_P$ and $Z_T$ of projectile and target, the reduced
mass number $A_{\rm{red}} = \frac{A_P A_T}{AP + A_T}$, and the plasma
temperature $T_9$ in $10^9$\,Kelvin. Although it has been shown that the
underlying simple approximation $S(E) \approx {\rm{const.}}$ does not hold
exactly \cite{Rau10}, the energy $E_0$ is still a reasonable estimate for
the most effective energy, and the energy window around $E_0$ is usually
called Gamow window.

According to Eq.~(\ref{eq:Ered}), the corresponding reduced energy
$E_{\rm{red},0}$ is given by 
\begin{equation}
E_{\rm{red},0} =  \frac{\bigl(A_P^{1/3}+A_T^{1/3}\bigr)}{Z_P Z_T} \, E_0
\approx 0.122\,{\rm{MeV}} \times T_9^{2/3} \times f(Z_P,Z_T,A_P,A_T)
\label{eq:E0red}
\end{equation}
where the function $f(Z_P,Z_T,A_P,A_T)$ is given by
\begin{equation}
f(Z_P,Z_T,A_P,A_T) = 
\left( \frac{A_P A_T}{Z_P Z_T} \right)^{1/3} \times
\frac{1+(A_P/A_T)^{1/3}}{[1+(A_P/A_T)]^{1/3}}
\label{eq:f}
\end{equation}
From the exponents of $1/3$ in Eq.~(\ref{eq:f}) already a relatively smooth
dependence of $E_{\rm{red},0}$ on the mass and charge numbers of projectile
and target can be expected. A closer inspection of Eq.~(\ref{eq:f}) shows that
there is a further compensation between the first and second factor in
Eq.~(\ref{eq:f}) which leads to a reduced Gamow energy $E_{\rm{red},0}$ which
is almost independent of target charge and mass for \al -induced reactions
($Z_P = 2$, $A_P = 4$). The first factor increases from light nuclei (with
$N_T \approx Z_T$ and thus $A_T/Z_T \approx 2$) towards heavier nuclei (with
$N_T \approx 1.5\,Z_T$ or $A_T/Z_T \approx 2.5$). The second factor decreases
with increasing target mass number $A_T$. Some numerical examples for
$f(Z_P,Z_T,A_P,A_T)$ are 
$f(^{20}{\rm{Ne}}) = 2.367$,   
$f(^{51}{\rm{V}}) = 2.288$, and  
$f(^{208}{\rm{Pb}}) = 2.165$. In the mass range under study $f$ varies only by
a few per cent, leading to
\begin{equation}
 E_{\rm{red},0} \approx 0.284\,{\rm{MeV}} \times T_9^{2/3}
\label{eq:E0redT}
\end{equation}
for the mass range under study. Explicitly, the numbers are $E_{\rm{red},0}
\approx 0.284$\,MeV for $T_9 = 1$,
$E_{\rm{red},0} \approx 0.451$\,MeV for $T_9 = 2$, and
$E_{\rm{red},0} \approx 0.591$\,MeV for $T_9 = 3$. 

\section{Update for $^{23}$Na + \al }
\label{sec:na23}
The total reaction cross section of $\alpha\ +$ $^{23}$Na is dominated by the
$^{23}$Na\rap $^{26}$Mg reaction at low energies. The $^{23}$Na\ran $^{26}$Al
reaction has a negative $Q$-value of about $-3$\,MeV.

Two new experimental data sets for the $^{23}$Na\rap $^{26}$Mg reaction have
been presented at this NPA-7 conference
\cite{Hub15,How15PRL,Tom15,Tom15PRL}. The results of both experiments agree 
with each other, but are in disagreement with the previous result of
\cite{Alm14}.

At the Aarhus accelerator forward kinematics was applied
\cite{Hub15,How15PRL}. The target composition of the NaCl target was carefully
monitored using Rutherford scattering on $^{23}$Na and
$^{\rm{nat}}$Cl. Because of the relatively low beam current, no significant
deterioration of the target was observed. Angular distributions for the $p_0$
and $p_1$ protons were measured over a broad angular range. Thus,
the derived absolute cross sections of the $^{23}$Na\rap $^{26}$Mg reaction
should be reliable within the given uncertainties in \cite{How15PRL} which are
of the order of $15 - 25$\,per cent.

At the ISAC facility at TRIUMF a helium-filled gas cell was mounted in the TUDA
scattering chamber for a measurement in inverse kinematics
\cite{Tom15,Tom15PRL}. The number of target nuclei was controlled via the gas
pressure, and the number of beam particles was measured by Rutherford
backscattering on the gas cell entrance window and an additionally mounted
thin gold foil. Also this procedure should lead to a reliable absolute
normalization of these experimental data. Absolute uncertainties of slightly
below 20\,\% at higher energies and about $30-40$\,\% at the lowest energies
are reported in \cite{Tom15PRL}.
\begin{figure}[ht]
\begin{minipage}{18pc}
\includegraphics[width=18pc]{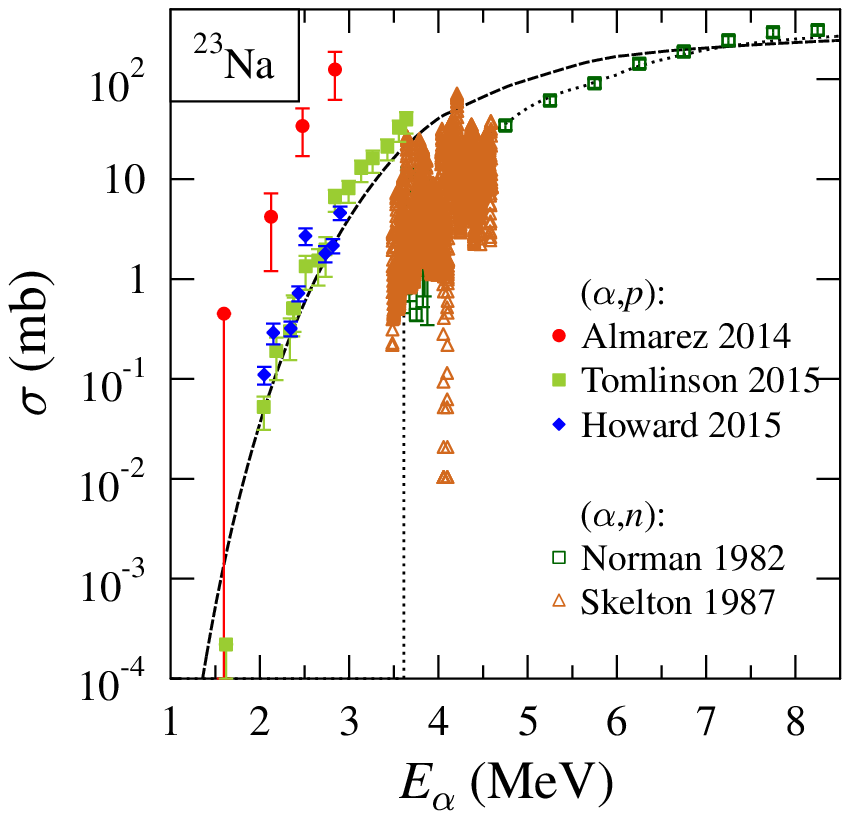}
\caption{
\label{fig:sigma}
Cross sections $\sigma$ of the $^{23}$Na\rap $^{26}$Mg and $^{23}$Na\ran
$^{26}$Al reactions {\it{vs.}}\ laboratory energy $E_{\rm{lab}}$. The
experimental data have been taken from
\cite{Alm14,How15PRL,Tom15PRL,Norman82,Skelton87}. The new data by Howard
{\it et al.}\ \cite{How15PRL} and Tomlinson {\it et al.}\ \cite{Tom15PRL}
are significantly lower than the earlier data by Almarez-Calderon {\it et
al.}\ \cite{Alm14}. The calculations are taken from
\cite{Mohr15}. Further discussion see text.
}
\end{minipage}\hspace{2pc}%
\begin{minipage}{18pc}
\includegraphics[width=18pc]{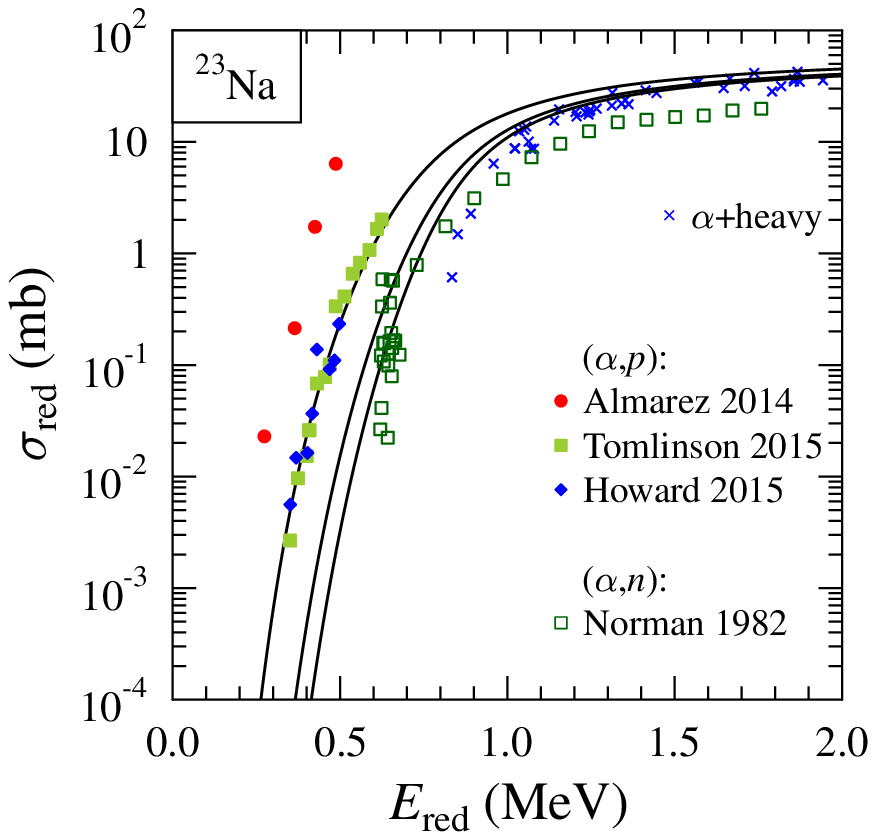}
\caption{
\label{fig:sred}
Reduced cross section \sred\ {\it{vs.}}\ reduced energy \Ered\ for the
$^{23}$Na\rap $^{26}$Mg reaction. The reduced cross sections \sred\ slightly
increase towards lower target masses; the expected range is indicated by three
calculations for $^{21}$Ne, $^{36}$Ar, and $^{51}$V (taken from
\cite{Mohr15}; for details see there).  The new data \cite{How15PRL,Tom15PRL}
fit into the general systematics whereas the earlier data \cite{Alm14} are
much higher.
}
\end{minipage} 
\end{figure}

In both experiments the target can be considered as relatively thick ($\approx
80-100$\,keV in the Aarhus experiment and $\approx 140$\,keV in the TRIUMF
experiment in the c.m.\ system). Thus, the data show a relatively smooth
energy dependence because of the averaging of the cross section in a
broad energy interval. The high data point of the Aarhus experiment
at $E_{\rm{lab}} \approx 2.5$\,MeV corresponds to a strong resonance which has
been seen also in earlier experiments \cite{Kup64,Whi74}. The new data are
shown together with previous data for the $^{23}$Na\rap $^{26}$Mg \cite{Alm14}
and $^{23}$Na\ran $^{26}$Al reactions \cite{Norman82,Skelton87} in
Fig.~\ref{fig:sigma} as cross sections {\it{vs.}}\ $E_{\rm{lab}}$ and in
Fig.~\ref{fig:sred} as \sred\ {\it{vs.}}\ \Ered . These figures are updates of
the corresponding Figs.~56 and 57 of the review \cite{Mohr15}.

It is obvious from Fig.~\ref{fig:sigma} that the new data from Howard {\it et
  al.}\ \cite{How15PRL} and Tomlinson {\it et al.}\ \cite{Tom15PRL} are in
good agreement with each other and with the prediction from the statistical
model (taken from \cite{Mohr15}). Contrary, the previous data by
Almarez-Calderon {\it et al.}\ \cite{Alm14} are more than one order of
magnitude higher. Fig.~\ref{fig:sred} shows the same data as reduced cross
sections \sred\ vs.\ reduced energies \Ered . Here it has been found in
\cite{Mohr15} that there is a smooth trend for increased \sred\ towards lower
target masses. The typical range of \sred\ for the $A \approx 20 - 50$ mass
range is indicated by three calculations for the total reaction cross sections
of $\alpha\ +$ $^{21}$Ne, $\alpha\ +$ $^{36}$Ar, and $\alpha\ +$ $^{51}$V (full
lines from left to right in Fig.\ \ref{fig:sred}; for details 
see \cite{Mohr15}). The new data of \cite{How15PRL,Tom15PRL} are located very
close to the expectation from systematics.

\section{Summary and Conclusions}
\label{sec:summ}
It has been shown in \cite{Mohr15} that \al -induced reaction cross sections
for targets in the $A \approx 20 - 50$ mass range show a relatively smooth and
systematic behavior. This behavior can be nicely visualized using so-called
reduced cross sections \sred\ and reduced energies \Ered . The present work
provides a simple relation between the reduced energy \Ered\ and the most
effective energy $E_0$ for astrophysical reaction rates (the so-called Gamow
window).

Four exceptions from the general systematics of \al -induced reaction cross
sections have been identified in
\cite{Mohr15}: $^{36}$Ar and $^{40}$Ar with smaller cross sections
\cite{Schwartz56}, $^{23}$Na \cite{Alm14} and $^{33}$S \cite{Bowers13} with
larger cross sections. New data \cite{How15PRL,Tom15PRL} for $^{23}$Na which
have been presented first at this NPA-7 conference \cite{Hub15,Tom15}
supersede the previous data \cite{Alm14}. Contrary to the previous data by
\cite{Alm14}, these new data \cite{How15PRL,Tom15PRL} indicate a regular
behavior of $^{23}$Na. New data for the remaining outliers $^{36}$Ar,
$^{40}$Ar (and $^{38}$Ar where no data are available) and $^{33}$S would be
helpful to confirm or reject the irregular behavior of these nuclei.

\smallskip
\noindent
{\underline{Note added:}} 
During review of this paper, an Erratum was published by Almarez-Calderon {\it
  et al.} \cite{Alm15}. The Erratum states that the $^{23}$Na\rap $^{26}$Mg
data of \cite{Alm14} have to be scaled down by a factor of 0.01, thus bringing
these data into agreement with the later data in \cite{How15PRL,Tom15PRL}.

\section*{Acknowledgments}
I thank J.\ Tomlinson and A.\ M.\ Howard for providing their data before final
publication. This work was supported by OTKA (K101328 and K108459).

\section*{References}

\end{document}